\documentclass[%
 reprint,
 amsmath,amssymb,
 aps,
]{revtex4-2}

\usepackage{graphicx}
\usepackage{dcolumn}
\usepackage{bm}
\usepackage[margin=20mm]{geometry}
\usepackage{amsfonts}
\usepackage{amssymb}
\usepackage{graphicx}
\usepackage{svg}
\usepackage{amsmath}
\usepackage{caption}
\usepackage[labelfont={small,bf}]{caption}
\usepackage[font={small}]{caption}
\usepackage{subcaption}
\usepackage{dsfont}
\usepackage{float}
\usepackage{setspace}
\usepackage{hyperref}
\usepackage{ragged2e}

\begin{document}

\preprint{APS/123-QED}

\title{Motility-induced phase separation drives aggregation in chemotactic systems}

\author{Letong Xu}
\author{Robert G. Endres}%
 \email{Email: r.endres@imperial.ac.uk}
\affiliation{%
 Department of Life Sciences \& Centre for Integrative Systems Biology and Bioinformatics, Imperial College London, London SW7 2AZ, United Kingdom }%

\date{\today}

\begin{abstract}
We present a unified continuum framework that combines chemotactic feedback with motility-induced phase separation (MIPS) to investigate the physical origin of aggregation in self-secreting cell populations. While classical Keller–Segel theory attributes pattern formation to chemotactic instabilities, we show that this mechanism breaks down in systems with proportional density-dependent (autocrine) production and uptake: linear stability analysis reveals that chemical fluctuations decouple from density, eliminating chemotactic aggregation at leading order. In contrast, introducing density-dependent motility restores instability through a purely mechanical mechanism corresponding to MIPS. Our results demonstrate that chemotactic feedback does not alter the nature of this instability, but instead enhances its growth rate and accelerates aggregation. This unifies two central paradigms of biological self-organization and revises the role of chemotaxis in collective cell migration.

\end{abstract}

\maketitle


\section*{Introduction}

Many biological systems exhibit self‐organized patterns that are both reproducible and robust despite environmental and intrinsic noise \cite{gierer1972theory,Karsenti2008,Needleman2017}. From the coordinated swarming of immune cells toward sites of infection to the aggregation of \textit{Dictyostelium discoideum} under starvation, such collective structures enable cooperative nutrient acquisition, coordinated migration, and developmental organization that cannot be achieved by individual cells alone \cite{kienle2021neutrophils,gregor2010onset,de2017critical}. Yet, despite experimental documentation of such patterns, the mechanistic principles that ensure their reproducible formation remain only partially understood.

\bigskip One of the most well‐studied mechanisms for collective cell migration is chemotaxis, in which cells detect and move along gradients of secreted signaling molecules, creating feedback between cell density and chemical concentration. A minimal mathematical description of chemotaxis is given by the Keller–Segel (KS) model \cite{KELLER1971225}, capturing a rich variety of instability behaviours (for recent reviews see \cite{arumugam2021keller}, Fig. 1A). In parallel, a growing area of interest in biophysics and microbiology is about how self-propelled particles undergo large-scale clustering without the need for explicit attractive forces. This phenomenon, known as motility-induced phase separation (MIPS), has been widely studied in theoretical models of active matter \cite{cates2015motility,caprini2020spontaneous,van2019interrupted,cates2013active}. It arises when particle motility decreases with increasing local density, producing a positive feedback loop that drives spontaneous separation into dense and dilute phases (Fig. 1B). Because active particles continuously consume energy to propel themselves, MIPS represents an intrinsically non-equilibrium mechanism of self-organization.

\bigskip While chemotaxis and MIPS are fundamental mechanisms of biological self-organization, they are typically studied in isolation despite being naturally coupled in living systems, leaving a significant gap in our theoretical understanding. Zhao {\it et al.} recently provided the first systematic framework that combines MIPS with chemotactic feedback (called CMIPS), revealing a complex stability landscape shaped by the competition between phase separation and signal-guided aggregation \cite{zhao2023chemotactic}. In their formulation, however, chemical dynamics include a constant supply term in space, so that signal production is independent of local cell density. In many biological systems, chemical production is instead density-dependent, as described in classical KS models. For example, in \textit{Dictyostelium discoideum}, cAMP secretion is amplified through positive feedback as cell density increases, reinforcing chemotactic aggregation \cite{de2017critical}. It is therefore essential to determine how density-dependent production modifies the stability structure of coupled MIPS–chemotaxis systems, and whether it merely shifts phase boundaries or qualitatively alters the underlying instability mechanism.

\bigskip Here, we focus on the role of quasi-realistic density-dependent production and degradation of chemoattractant and analyze how this modification reshapes the stability structure of the system. We then perform linear stability analysis and parameter-matched simulations to compare the self-production model directly with the standard KS formulation. This framework clarifies how density-dependent production modifies the interplay between chemotaxis and MIPS that underlies collective pattern formation. Our results demonstrate that, in this class of self-secreting systems, aggregation is initiated by mechanical effects rather than chemical feedback, with chemotaxis acting as a secondary amplification mechanism. This has fundamental implications for many phenomena in biology, ranging from microbial communities to embryonic and tumor development.

\begin{figure}
    \centering  \includegraphics[width=\linewidth]{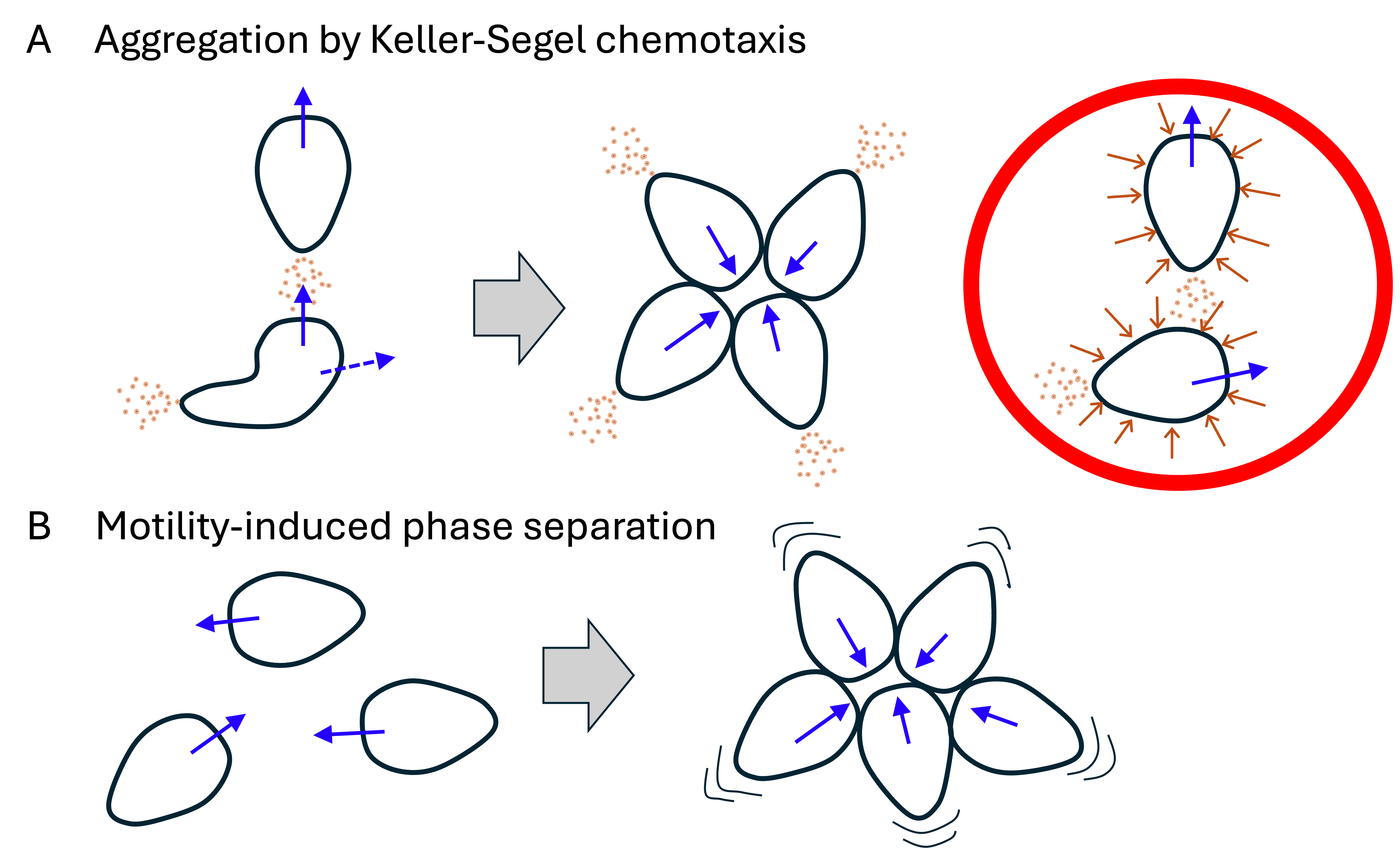}
    \caption{\justifying
    {\bf Two modes of aggregation: chemically versus mechanically driven.} (A) Aggregation by Keller-Segel (KS) type chemotaxis, based on secretion of attractant (e.g. at the rear of the cell), sensing, and chemotaxis. (red-circle inset) This mechanism, however, cannot initiate aggregation for quasi-realistic density-dependent uptake (beige arrows). (B) Jamming by motility-induced phase separation (MIPS). Blue arrows indicate direction of motion. Dashed arrow indicates initial direction before sensing and turning.}
    \label{fig:schematic}
\end{figure}

\section*{Unified continuum model}

We introduce a unified continuum description in which chemotaxis and MIPS arise as limiting cases. We consider the evolution of cell density
$\phi(\mathbf{x},t)$ and chemoattractant concentration $c(\mathbf{x},t)$:

\begin{align}
\partial_t \phi &= \nabla \cdot \Big[ M(\phi)\nabla \mu(\phi)
- \chi_0\, \phi \nabla c \Big], \label{eq:phi_master} \\
\partial_t c &= D_c \nabla^2 c + a\, \phi - k(\phi)\,c.
\label{eq:c_master}
\end{align}

Here, $M(\phi)$ denotes the mobility and $\mu(\phi)$ an effective chemical
potential encoding density-dependent motility, while $\chi_0$ is the chemotactic
sensitivity. The chemical field diffuses with coefficient $D_c$, is produced at
rate $a\phi$, and degrades at rate $k(\phi)c$, which may depend on the local cell density. The choice $k(\phi)=k$ corresponds to minimal signaling with density-independent degradation \cite{zhao2023chemotactic}, whereas $k(\phi)=k\phi$ represents autocrine signaling with uptake proportional to the local cell density. This formulation unifies several previously studied models within a common framework. \\

In particular, the standard KS form considered below is recovered for constant mobility $M(\phi)=D_\phi$ and linear
$\mu(\phi)$, although numerous generalizations incorporate nonlinear diffusion, density-dependent chemotactic sensitivity, volume exclusion, and alternative signaling kinetics \cite{PainterHillen2002,
HillenPainter2009}.
Here, however, density-dependent motility can itself generate a mechanical instability through MIPS, providing a route to aggregation that is independent of the chemotactic instability.
Specifically, nonlinear $\mu(\phi)=\mu_h(\phi)-\kappa\nabla^2\phi$ captures MIPS-type transport through
density-dependent slowing. Within this unified framework, the system is controlled by two independent mechanisms: (i) the transport law governing how cells move (diffusive versus MIPS-type), and (ii) the nature of the signaling feedback (minimal versus autocrine). All models analyzed in the following are obtained as limiting cases of
Eqs.~\eqref{eq:phi_master}--\eqref{eq:c_master}. \\

The unified framework also clarifies the origin of effective diffusion in the MIPS sector. Considering the diffusion-like term $\nabla\cdot(M_0\phi\nabla\mu_h)$ near the uniform state $\phi_0$, we write $\phi=\phi_0+\delta\phi$ and expand $\mu_h(\phi)\approx\mu_h(\phi_0)+\mu_h'(\phi_0)\delta\phi$. Keeping only linear terms gives $\nabla\cdot(M_0\phi\nabla\mu_h)\approx M_0\phi_0\mu_h'(\phi_0)\nabla^2\delta\phi$ (see Supplemental Material \cite{SupplementalMaterial} for details). Thus, for small perturbations, the CMIPS dynamics reduces to an effective diffusion with coefficient $D_\phi^{\mathrm{eff}} = M_0\phi_0\mu_h'(\phi_0)$, which may become negative when $\mu_h'(\phi_0)<0$, indicating aggregation. In the linear regime, the $q^2$-slope of the CMIPS growth rate (with $q$ the wavenumber) is therefore $D_\phi^{\mathrm{eff}}$, whereas the KS model has slope $D_\phi$.

\bigskip To compare the different limits of the unified continuum framework on equal footing, we introduce nondimensional variables separately for the MIPS-based and diffusive transport classes (see Supplemental Material for details). The dimensionless parameters are P\'eclet number $\mathrm{Pe}_C=\chi_0/M_0$ for the MIPS class and $\mathrm{Pe}_C=\chi_0/D_\phi$ for the diffusive class, together with $\alpha_0=M_0/D_c$ (or $D_\phi/D_c$ in the diffusive class of KS models) and Damk\"ohler number $\mathrm{Da}_0=\kappa k/D_c$. These nondimensional forms are used to compare phase diagrams, dispersion relations, and numerical simulations across the different limiting cases, while all physical conclusions are independent of the specific choice of scaling.

\begin{figure*}
    \centering
    \includegraphics[width=1\linewidth]{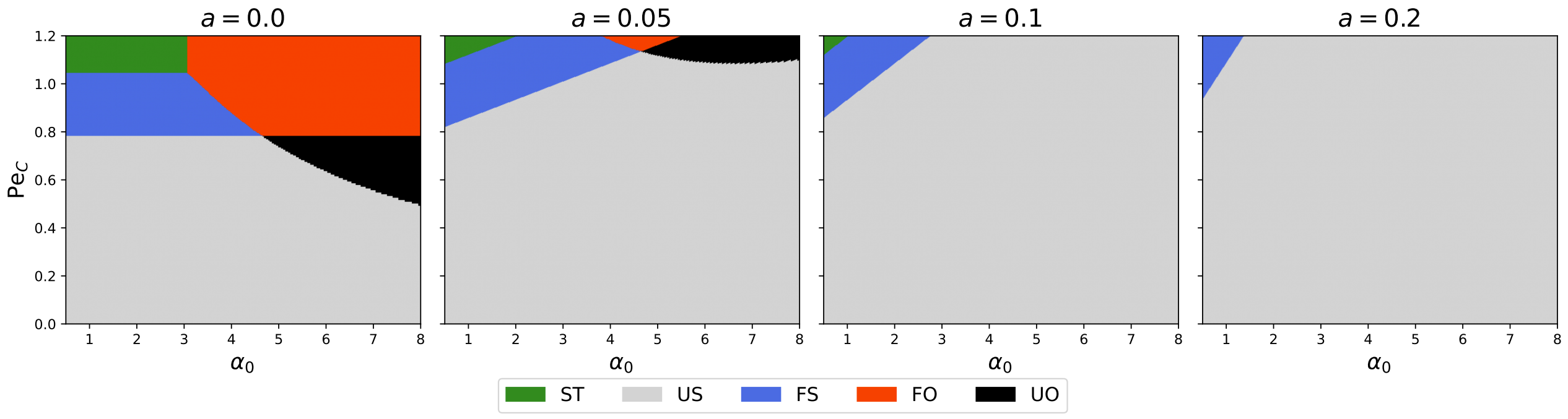}
    \caption{\justifying
    {\bf Evolution of the phase diagram as the strength of density-dependent production $a$ is increased in the source term $S + a\phi$.} Each panel shows the linear stability classification in the $(\alpha_0, \mathrm{Pe}_C)$ parameter space for increasing values of $a$. The homogeneous state (ST, green) is stable at high $\mathrm{Pe}_C$. Stationary regimes include unbounded stationary (US, grey) and finite-wavelength stationary (FS, blue), while oscillatory regimes include finite-wavelength oscillatory (FO, red) and uniform oscillatory (UO, black).
}

    \label{fig:phasediagram}
\end{figure*}

\section*{Density-dependent production reshapes stability}

To determine the role of density-dependent chemical production, we compare the linear stability of systems with a constant source term $S$ and with a self-production term of the form $S + a\phi$. The resulting phase diagrams for increasing values of $a$ are shown in Fig.~\ref{fig:phasediagram}. For $a = 0$, the system exhibits a rich stability structure, corresponding to the model in \cite{zhao2023chemotactic}. As the chemotactic strength $\mathrm{Pe}_C$ increases, the homogeneous state becomes unstable and transitions into stationary regimes (US, FS) as well as oscillatory instabilities (FO, UO; see figure caption for details). These oscillatory regimes occupy a substantial region of parameter space.\\

Introducing self-production through $S + a\phi$ hence leads to a restructuring of the phase diagram. Even for modest values of $a$, oscillatory regimes shrink and shift to larger values of $\alpha_0$ and $\mathrm{Pe}_C$. As $a$ increases further, oscillatory instabilities are strongly suppressed and nearly eliminated, while stationary regimes expand and dominate the parameter space. These results show that self-production qualitatively reshapes the instability structure of the system, strongly suppressing oscillatory regimes and altering the feedback between density and chemical signaling. This highlights the need to incorporate density-dependent production into the theoretical framework explicitly.

\section*{Linear instability and aggregation mechanism}
\begin{figure*}
    \centering
    \includegraphics[width=0.95\linewidth]{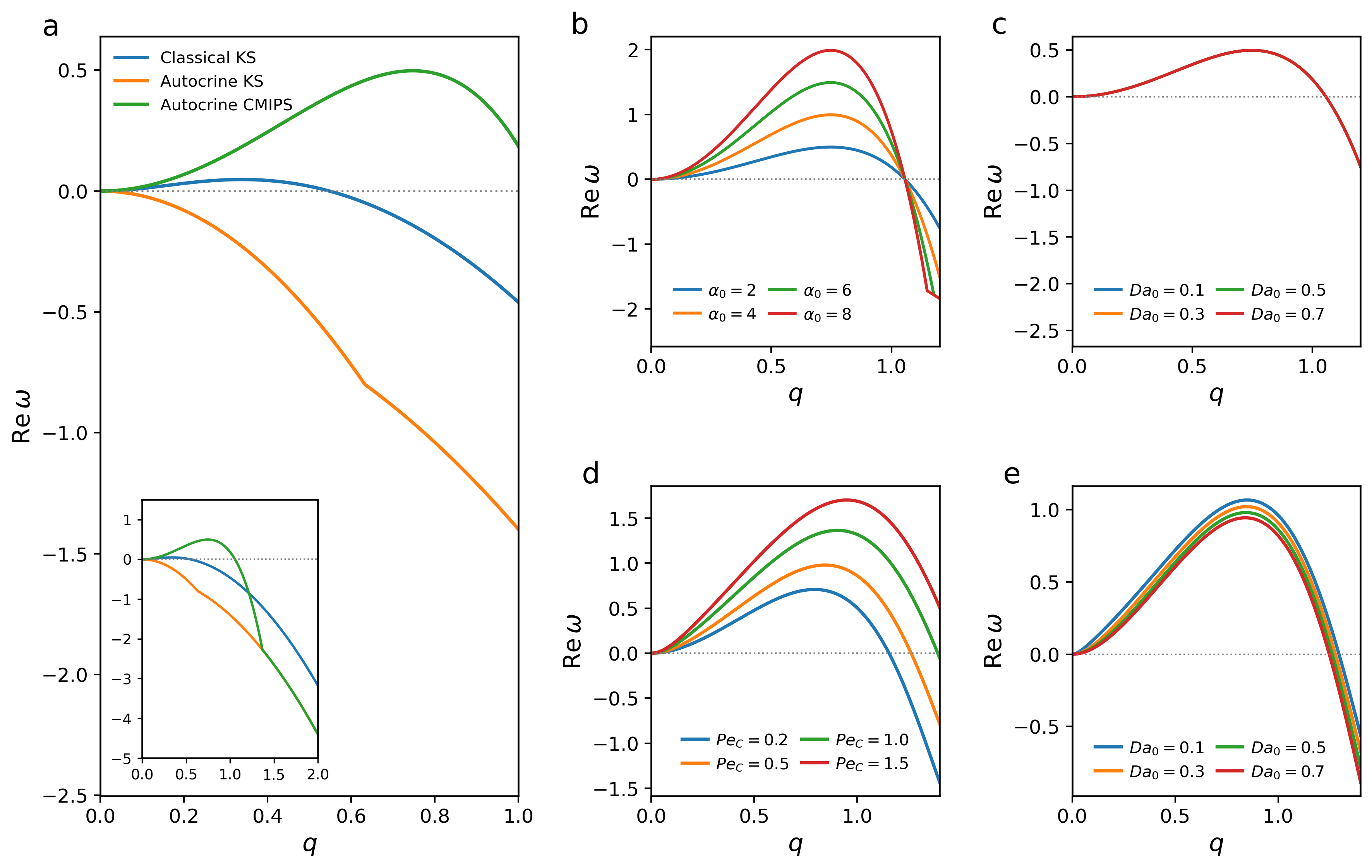}
    \caption{\justifying
    \textbf{Dispersion relations illustrating loss, restoration, and amplification of instability.} (a) Representative growth rates $\omega(q)$ for three limits of the unified where $q$ is nondimensional wavenumber. The classical KS model exhibits a finite band of unstable modes. Autocrine signaling suppresses this chemotactic instability, whereas MIPS restores a long-wavelength mechanical instability with a finite fastest-growing wavenumber. (b) Autocrine CMIPS with varying mobility ratio $\alpha_0$ at fixed $Da_0=0.5$. Increasing $\alpha_0$ raises the growth rate while leaving the selected wavelength nearly unchanged. (c) Autocrine CMIPS with varying $Da_0$ at fixed $\alpha_0=2$. The dispersion curves nearly collapse, indicating weak sensitivity to uptake kinetics. (d) Minimal CMIPS with varying chemotactic strength $\mathrm{Pe}_C$ at fixed $\alpha_0=2$ and $Da_0=0.5$. Larger $\mathrm{Pe}_C$ enhances the peak growth rate and modestly shifts the unstable band. (e) Minimal CMIPS with varying $Da_0$ at fixed $\alpha_0=2$ and $\mathrm{Pe}_C=0.5$. Increasing $Da_0$ lowers the peak growth rate as faster signal decay weakens chemotactic reinforcement.}
    \label{fig:dispersionT}
\end{figure*}

To identify the mechanism that initiates aggregation in each limiting case, we analyse the stability of the homogeneous state 
$\phi=\phi_0$ and $c=c_0$ by performing linear stability analysis.
Perturbing the steady state with small fluctuations 
$\delta\phi, \delta c \propto e^{iqx+\omega t}$ 
yields the dispersion relation $\omega(q)$ (see Supplemental Material for the derivation) and the dispersion relation diagrams (Fig.~\ref{fig:dispersionT}) to be discussed below.

\subsection{Autocrine signaling prevents Keller–Segel pattern formation}

We first consider the classical KS limit within the unified continuum framework Eqs. \ref{eq:phi_master} and \ref{eq:c_master}. For diffusive transport and minimal signaling, corresponding to constant mobility $M(\phi)=D_\phi$ and density-independent degradation $k(\phi)=k$, linear stability analysis of the homogeneous state yields the standard chemotactic instability. The resulting dispersion relation exhibits a finite band of unstable wavenumbers, reflecting the positive feedback between density fluctuations and self-generated chemical gradients (Fig.~\ref{fig:dispersionT}a).

\bigskip We next consider diffusive transport with autocrine signaling, obtained in the unified framework by taking $M(\phi)=D_\phi$ and $k(\phi)=k\phi$, such that
\begin{align}
\partial_t \phi &= \nabla \cdot \left[D_\phi \nabla \phi - \chi_0 \phi \nabla c \right], \\
\partial_t c &= D_c \nabla^2 c + a \phi - k \phi c.
\end{align}

 Linear stability analysis of the homogeneous state reveals that the coupling between density and chemoattractant fluctuations vanishes at linear order (see Fig. S1 and Supplemental Material for details). At steady state $(\phi_0,c_0)$ the balance condition $k\phi_0 c_0 = a\phi_0$ implies $kc_0=a$. Linearizing about this state yields $\omega\,\delta c = -D_c q^2 \delta c - k\phi_0 \delta c$, so that $\delta c$ evolves independently of the density perturbation $\delta\phi$ at first order. Consequently, the dispersion relation factorizes into independent density and chemical modes. Without chemotactic feedback, density perturbations relax diffusively, and no linear instability occurs, so the autocrine KS system cannot generate patterns at the linear level  (Fig.~\ref{fig:dispersionT}a). This suppression of chemotactic feedback explains the absence of aggregation in the autocrine KS system at early times, and highlights that chemotaxis alone is insufficient to generate clustering in this regime.

\begin{figure}
    \centering
    \includegraphics[width=1\linewidth]{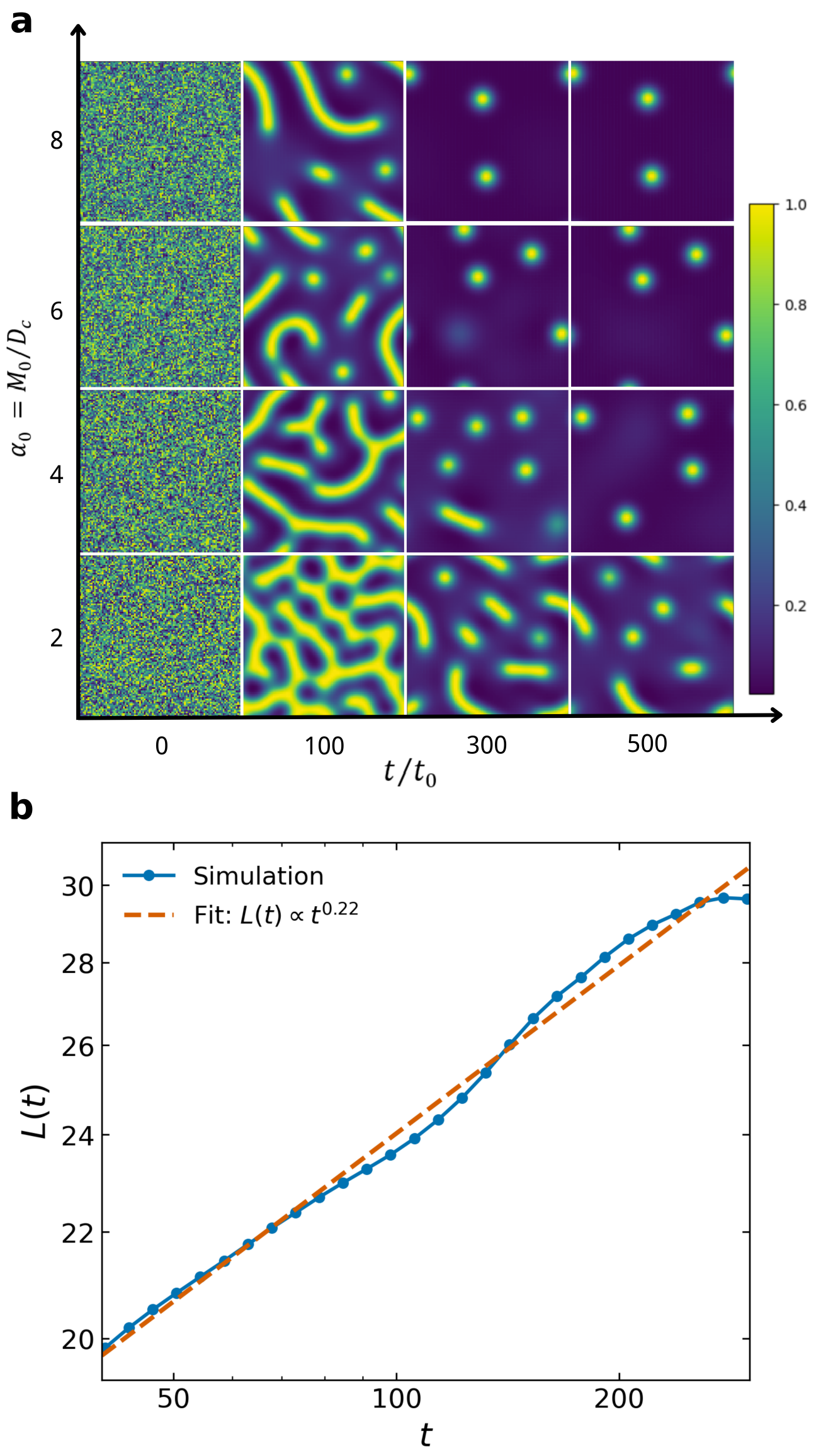}
    \caption{\justifying
\textbf{Time evolution and coarsening in the autocrine CMIPS model.}
\textbf{(a)} Rows correspond to increasing mobility ratio $\alpha_0=2,4,6,8$ (bottom to top), while columns show successive times $t/t_0=0,100,300,500$ (left to right). For all parameter sets, the system follows the same qualitative sequence: amplification of initial density fluctuations, coarsening driven by MIPS, and formation of discrete high-density droplets. No travelling or oscillatory patterns are observed.
\textbf{(b)} Characteristic domain size $L(t)$ during the intermediate coarsening regime for a representative parameter set with $\alpha_0=5$, $\mathrm{Pe}_C=0.5$, and $\mathrm{Da}_0=0.5$.
Here, $L(t)=2\pi/\langle q\rangle$ measures the typical spatial size of the density domains, where $\langle q\rangle$ is the power-weighted mean wavenumber of the density fluctuations (see Supplemental Material for details). The blue curve shows simulation data, and the orange dashed line shows a power-law fit, $L(t)\propto t^{0.22}$, over $40\leq t\leq300$. The fitted exponent characterises growth within this time interval. Other parameters for (b) are $\phi_0=0.8$, $\phi_m=0.9$, $\kappa=5$, and $\mathrm{Pe}_R=10^{-3}$. The full coarsening trajectory and analysis are provided in Supplemental Material Fig.~S3b.}
    \label{fig:simulation}
\end{figure}

\subsection{Early aggregation in a self-secreting system is mechanically driven}

We now introduce MIPS transport term
while retaining the autocrine signaling dynamics (autocrine CMIPS):
\begin{align*}
\partial_t \phi
&= \nabla\!\cdot
\left[
M_0 \phi \nabla \big(\mu_h(\phi)-\kappa\nabla^2\phi\big)
-\chi_0 \phi \nabla c
\right], \\
\partial_t c
&= D_c \nabla^2 c + a\phi - k\phi c.
\end{align*}
As shown above, autocrine signaling eliminates the linear chemotactic coupling, so that the chemical field does not contribute to the instability at first order. Any aggregation must therefore originate from the transport dynamics. The dispersion relation confirms that the homogeneous state becomes unstable through the MIPS mechanism (Fig.~\ref{fig:dispersionT}a). The resulting dispersion relation contains only real eigenvalues, $\mathcal{M}(q)=-M_0\phi_0\,q^2\big(\partial_\phi\mu_h+q^2\big)$ and $\mathcal{D}(q)=-D_c q^2-k\phi_0$, ruling out oscillatory or traveling modes. All unstable spectra are U-type, with unstable modes extending to arbitrarily small nonzero \(q\), a finite fastest-growing mode, and suppression at large \(q\). This structure is characteristic of a coarsening instability.\\

Varying the mobility ratio $\alpha_0 = M_0/D_c$ modifies the strength of the mechanical (MIPS) contribution relative to chemical diffusion. Increasing $\alpha_0$ raises the peak growth rate but leaves the selected wavenumber unchanged (Fig.~\ref{fig:dispersionT}b). Thus, faster-moving cells do not alter the characteristic length scale of the initially unstable mode; instead, they amplify it more rapidly. Consistent with this prediction, numerical simulations (Fig.~\ref{fig:simulation}a) show that populations with larger $\alpha_0$ nucleate clusters earlier and coarsen more quickly, while the spacing between emerging clusters remains essentially unchanged. \\

Likewise, at fixed $\alpha_0$, varying the uptake rate $k$ (and hence $Da_0$) does not affect the dispersion spectrum (Fig.~\ref{fig:dispersionT}c). All real-part curves collapse onto a single curve, while the imaginary parts remain identically zero. This insensitivity arises because the mechanical eigenvalue $\mathcal{M}(q)$ is independent of the uptake rate, whereas the chemical branch $\mathcal{D}(q)$ is strictly stabilizing. As a result, MIPS, rather than chemical transport or uptake kinetics, controls initial nucleation, wavelength selection, and early growth of density clusters in a self-secreting population.

\subsection{Chemotactic feedback enhances MIPS instability}

We next consider the MIPS transport dynamics coupled to minimal signaling (minimal CMIPS), in which the system obeys
\begin{align*}
\partial_t \phi
&= \nabla\!\cdot
\left[
M_0 \phi \nabla \big(\mu_h(\phi)-\kappa\nabla^2\phi\big)
-\chi_0 \phi \nabla c
\right], \\
\partial_t c
&= D_c \nabla^2 c + a\phi - k c.
\end{align*}

Here, the density and chemical fields are linearly coupled. Linearizing the equations shows that density fluctuations are driven by the chemotactic drift term $\chi_0 \phi_0 q^2 \delta c$, while the chemical field responds to density perturbations through the production term $a\,\delta\phi$. These two processes generate a positive feedback loop: local density increases raise the chemoattractant concentration, which biases further motion toward dense regions. Since the production rate $a$ only sets the overall strength of this feedback, we fix $a=1$ and focus on the remaining control parameters below.\\

\begin{figure}
    \centering
    \includegraphics[width=0.99\linewidth]{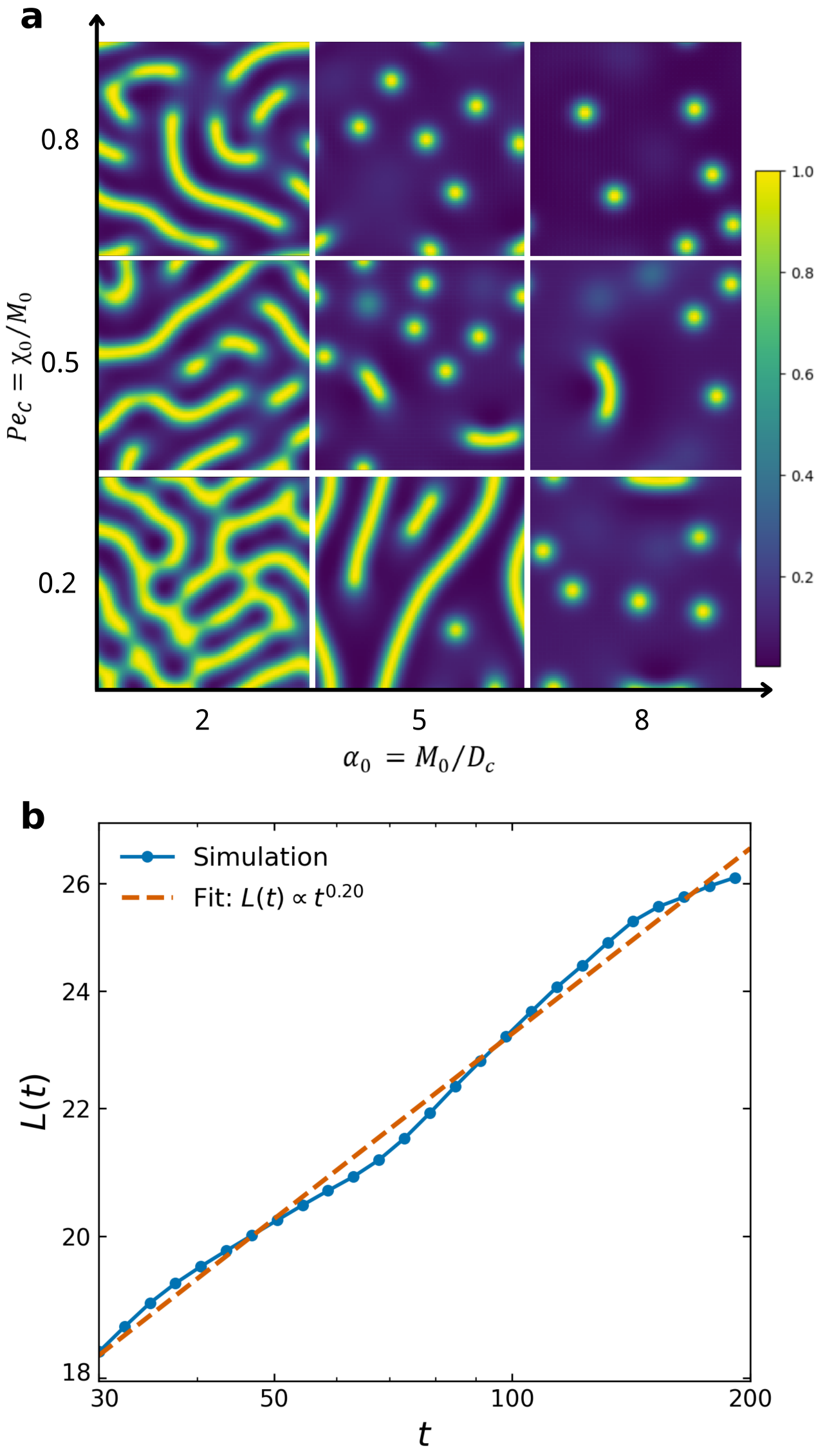}
    \caption{\justifying
\textbf{Pattern formation and coarsening in the minimal CMIPS model.} \textbf{(a)} Rows correspond to increasing chemotactic strength $\mathrm{Pe}_C=0.2,0.5,0.8$ (bottom to top), while columns show increasing mobility ratio $\alpha_0=2,5,8$ (left to right). Snapshots are taken at a fixed time $t=150$. Increasing $\mathrm{Pe}_C$ produces more pronounced aggregation at this time, consistent with the enhanced growth rate predicted by the dispersion relation. Larger $\alpha_0$ also promotes cluster formation. No travelling or oscillatory patterns are observed, while the density domains continue to coarsen.
\textbf{(b)} Characteristic domain size $L(t)$ during the intermediate coarsening regime for $\alpha_0=5$ and $\mathrm{Pe}_C=0.5$, corresponding to the central parameter combination in (a), with $\mathrm{Da}_0=0.5$. The blue curve shows simulation data, and the orange dashed line shows a power-law fit, $L(t)\propto t^{0.20}$, over $30\leq t\leq200$. The fitted exponent characterizes growth within this time interval. Other parameters for (b) are $\phi_0=0.8$, $\phi_m=0.9$, $\kappa=5$, and $\mathrm{Pe}_R=10^{-3}$. The full coarsening trajectory and analysis are provided in Supplemental Material Fig.~S2b.}
    \label{fig:simulation2}
\end{figure}

The strength of chemotactic feedback is controlled by $Pe_C$, which measures the relative strength of chemotactic drift compared with particle motility. Increasing $Pe_C$ therefore corresponds to a stronger chemotactic response to chemical gradients. As shown in Fig.~\ref{fig:dispersionT}d, larger $Pe_C$ substantially enhances the peak growth rate of the dispersion relation and slightly shifts the most unstable mode to larger wavenumbers, indicating faster amplification of density fluctuations and thus more rapid aggregation. This trend is confirmed by numerical simulations (Fig.~\ref{fig:simulation2}a), where increasing $Pe_C$ leads to earlier formation of dense clusters and more pronounced coarsening.\\

In the autocrine model discussed above, the dispersion relation is completely determined by the MIPS transport dynamics. When minimal signaling is introduced, the chemotactic coupling remains active at linear order, leading to a substantially larger peak growth rate. However, the spectrum still exhibits the characteristic MIPS structure: the growth rate is positive over a finite band of long wavelengths, reaches a single maximum, and decays at larger wavenumbers, while the imaginary part remains identically zero. \\

The impact of chemotactic feedback is clearly seen in the dynamics. In the minimal-signaling model, well-defined clusters emerge rapidly (Fig.~\ref{fig:simulation2}a), with clear structures already visible at $t \approx 150$. In contrast, the autocrine model requires significantly longer times ($t \approx 500$, Fig.~\ref{fig:simulation}a) to reach comparable levels of aggregation. This separation of timescales suggests that chemotaxis primarily acts by accelerating the underlying MIPS instability, rather than changing its qualitative mechanism: the patterns remain stationary, but density fluctuations amplify and aggregate much earlier. Following initial pattern formation, both signaling regimes exhibit coarsening (see Figs. ~\ref{fig:simulation}b and ~\ref{fig:simulation2}b), with the characteristic domain size growing approximately as \(L(t)\sim t^{0.2}\) over the fitted time windows in representative simulations (see Supplemental Material, Figs. S2 and S3, for details).

\section*{Chemical signaling transmits MIPS-generated patterns between populations}

Having established that MIPS can provide the pattern-forming instability independently of chemotactic feedback, we next asked whether a MIPS-generated pattern could organize a second population that is stable in isolation. We therefore considered a two-population system consisting of a stable diffusive population, $\phi_1$, and a CMIPS-unstable population, $\phi_2$. The stable population corresponds to the autocrine KS regime considered above, whereas the second population provides the MIPS-driven instability.

\bigskip
We considered two versions of this two-population system, based on the two CMIPS models studied above: autocrine CMIPS and minimal CMIPS. In both cases, the populations interact through a shared chemical field, but the signaling architectures differ rather than representing a parameter-matched change in uptake alone. Under autocrine signaling, both populations contribute proportionally to chemical production and uptake, whereas under minimal signaling the CMIPS population provides the density-dependent source while degradation remains density independent. This distinction determines whether the MIPS-generated density pattern is converted into a spatial chemical cue that can be sensed by the stable population.

\begin{figure}
    \includegraphics[width=1\linewidth]{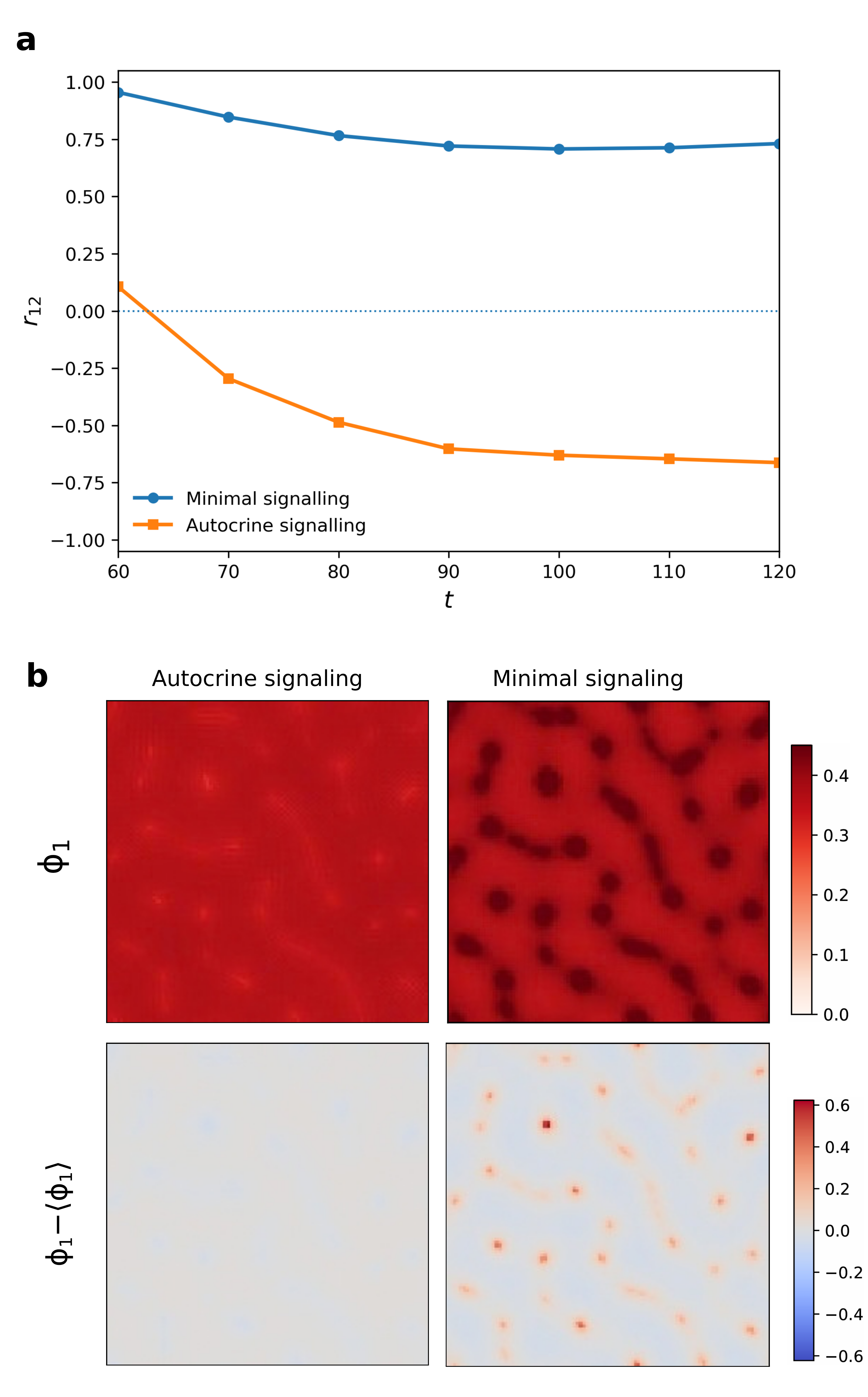}
    
    \caption{\justifying
    \textbf{Chemical signaling controls transmission of MIPS-generated spatial organisation.} 
    \textbf{(a)} Spatial Pearson correlation, \(r_{12}\), between the stable KS population \(\phi_1\) and the CMIPS population \(\phi_2\) under autocrine and minimal signaling over \(60 \leq t \leq 120\). Minimal signaling system maintains a strong positive correlation between the populations, whereas the autocrine signaling system yields a negative correlation.
    \textbf{(b)} Stable-population density, $\phi_1$, and density fluctuation, $\phi_1-\langle\phi_1\rangle$, at $t=120$. The stable population remains nearly homogeneous in the autocrine system but develops spatial structure under minimal signaling. The same color scale is used between signaling conditions within each row.}
    \label{fig:two_population_transmission}
\end{figure}

\bigskip
For autocrine signaling, production and uptake scale with the same local density,
\begin{equation}
\frac{\partial c}{\partial t}
=
D_c\nabla^2c
+
a(\phi_1+\phi_2)
-
k(\phi_1+\phi_2)c.
\end{equation}
At $c=a/k$, production and uptake cancel locally, so density heterogeneity does not generate a persistent chemical gradient. Consequently, although the CMIPS population develops spatial structure, the stable population remains nearly homogeneous (Fig.~\ref{fig:two_population_transmission}). 

\bigskip
By contrast, under minimal signaling,
\begin{equation}
\frac{\partial c}{\partial t}
=
D_c\nabla^2c
+
a_2\phi_2
-
kc,
\end{equation}
the density pattern of the CMIPS population generates a heterogeneous chemical field. This recruits the stable population into the emerging structure. Pearson correlation was used to quantify spatial co-localization between $\phi_1$ and $\phi_2$. It remains strongly positive under minimal signaling, whereas it becomes negative in the autocrine system (Fig.~\ref{fig:two_population_transmission}a). Spatial maps at $t=120$ show the same contrast: $\phi_1$ remains nearly uniform in the autocrine case but develops clear spatial structure under minimal signaling (Fig.~\ref{fig:two_population_transmission}b). The fluctuation field $\delta \phi_1 = \phi_1 - \langle \phi_1 \rangle$
further highlights the weak response in the autocrine case and the strong spatial modulation produced by minimal signaling. These results separate pattern generation from pattern transmission. MIPS generates spatial structure, while transmission to an otherwise stable population requires a shared chemical field that remains spatially responsive to the MIPS-generated density pattern. Full time-resolved simulations are shown in Supplemental Material Figs.~S4 and S5.

\section*{Discussion}
Understanding collective aggregation requires distinguishing the mechanism that initiates a pattern from the processes that reinforce it. The unified framework developed here makes this distinction explicit by considering transport and signaling as independent components of the dynamics (Table~\ref{tab:framework}). Under the proportional production–uptake kinetics studied here, autocrine signaling suppresses the classical chemotactic instability, while density-dependent motility slowdown retains a mechanical route to aggregation through MIPS. With minimal signaling, chemotaxis reinforces the growth of this mechanical instability. Our heterogeneous-population simulations extend this picture further, showing how chemical communication can recruit an otherwise stable population into a structure generated by another.

\begin{table*}[htbp]
\centering
\renewcommand{\arraystretch}{1.4}
\setlength{\tabcolsep}{12pt}
\begin{tabular}{c|c|c}
\hline
\textbf{Transport}
 & \textbf{Minimal signaling} ($k(\phi)=k$) 
 & \textbf{Autocrine signaling} ($k(\phi)=k\phi$) \\
\hline
\textbf{Diffusive} 
& Classical KS instability 
& No linear instability (decoupling) \\
\hline
\textbf{MIPS} 
& Chemotaxis-enhanced MIPS aggregation 
& MIPS-driven aggregation (mechanical) \\
\hline
\end{tabular}
\caption{\justifying {{\bf Summary of aggregation mechanisms in the unified framework.} The system is
organized along two axes: transport (KS-like diffusive vs.\ MIPS) and signaling
(minimal vs.\ autocrine). Classical chemotactic instability is recovered only
for diffusive transport with minimal signaling. Autocrine signaling suppresses
this instability at linear order, while MIPS-type transport restores aggregation
through a mechanical mechanism. Chemotaxis enhances but does not create the
instability in the MIPS regime.}
}
\label{tab:framework}
\end{table*}

\bigskip  
Classical KS theory attributes aggregation to positive feedback between local cell density and chemoattractant accumulation \cite{KELLER1971225,keller1970initiation,keller1970conflict}. Our results show why this feedback depends on how the signal is both produced and removed. When production and uptake increase proportionally with density, their contributions to the chemical response cancel at linear order: more cells produce more signal, but also remove it more rapidly. A density fluctuation therefore need not generate the chemical gradient required to recruit additional cells. MIPS provides a route to aggregation that does not require this feedback, because density-dependent slowing can destabilize the population directly, consistent with active matter studies of phase separation \cite{cates2015motility,fily2012athermal}.
Under minimal signaling, density fluctuations generate chemical gradients, allowing chemotaxis to amplify their growth in the MIPS-unstable regime examined here. Taken together, these results identify MIPS as the underlying instability in this regime, with chemotaxis acting as an amplifier of mechanically initiated clustering. This contrasts with the classical view that chemotaxis is the primary driver of aggregation, suggesting that mechanical effects, such as crowding-induced reduction in motility, may play a dominant role. \\

This conclusion resonates with previous experimental observations by Peruani {\it et al.} in \textit{Myxococcus xanthus} system, where clustering and coarsening can arise even in the absence of chemotactic signaling, driven instead by self-propulsion and volume exclusion \cite{peruani2012collective}. More broadly, a growing body of work shows that bacterial behavior is highly sensitive to mechanical cues, including surface adhesion, fluid shear, and spatial confinement \cite{dufrene2020mechanomicrobiology,marra2025microfluidics,palalay2023shear}. These mechanical inputs regulate motility, attachment, and biofilm formation, indicating that physical interactions can play a central role in collective organization. Together, these findings suggest that aggregation in active biological systems may be fundamentally mechanical in origin, with MIPS providing the primary instability mechanism and chemotaxis acting as a secondary, modulatory effect.\\

The heterogeneous-population simulations reveal an additional role for chemical signaling. Beyond accelerating MIPS-driven aggregation within a population, a spatially varying chemical field can transmit mechanically generated organization between populations. The comparison between autocrine and minimal signaling therefore shows that mechanical clustering alone is insufficient to organize a second population; the chemical signal must preserve information about the density pattern. Thus, not every population must be independently unstable: one population may generate the pattern through MIPS, while chemical signaling recruits otherwise stable populations into the emerging structure.\\

This prediction on one population entraining another could be tested directly in mixed-population experiments in {\it D. discoideum}. A signaling-competent population could be mixed with fluorescently labelled cells that are defective in autonomous cAMP relay but retain chemotactic responsiveness. The model predicts recruitment of the latter into aggregates only when density patterns in the signaling population generate spatially heterogeneous cAMP fields; this could be quantified experimentally using the same population cross-correlation and aggregate-enrichment measures employed here. There is encouraging precedent for exactly this type of experiment. 
For example, {\it Dictyostelium} cells defective in cAMP relay have been reported to co-aggregate with wild-type cells \cite{Matsuoka2004}. And acaA- cells, which lack the aggregation-stage adenylyl cyclase required for cAMP relay, fail to aggregate autonomously but can participate in development when mixed with wild-type cells \cite{Pitt1993,Soderbom1999}.\\

The present framework has several limitations. The cancellation of chemotactic feedback in the autocrine limit relies on production and uptake having the same proportional dependence on cell density, and should therefore not be generalized to all self-secreting populations. Nonlinear secretion, saturating uptake, or adaptive receptor responses could restore or modify density–chemical feedback (see Supplemental Material). The balance between mechanical and chemotactic mechanisms may also depend on the physical environment: confinement, substrate friction, hydrodynamic interactions, and population variability could modify transport and aggregation dynamics \cite{aranson2022bacterial,xu2023geometrical,barriuso2022simulating}, along with dynamical noise and variability within each population \cite{cross1993pattern,mitchell2018identifying}. Experimental studies in systems such as {\it Myxococcus xanthus} or {\it Dictyostelium discoideum} could test whether early aggregation is mechanically initiated and subsequently amplified by chemotactic signaling \cite{de2017critical}.\\

In conclusion, we have developed a unified continuum framework that clarifies how transport and signaling kinetics jointly control the onset of collective aggregation. Our results show that autocrine signaling can suppress the classical chemotactic instability, while MIPS-type transport restores robust pattern formation via a mechanical route, with chemotaxis amplifying growth under minimal signaling. Chemical signaling instead can transmit MIPS-generated spatial organization to an otherwise stable population, extending mechanically driven aggregation across populations. These findings provide a revised mechanism of clustering in autocrine populations and suggest that mechanical interactions may play a broader role in biological self-organization than is often assumed. More generally, the framework offers a flexible basis for future studies of coupled active matter and signaling dynamics in microbial, cellular, and biomedical systems.

\bigskip
\bigskip
\small
\subsection*{Numerical Methods}
Numerical simulations were performed using a common nondimensionalization to compare the transport and signaling regimes within the unified framework. Length and time were scaled by $l_0=\sqrt{\kappa}$ and $t_0=\kappa/D_c$, respectively, and the chemoattractant concentration by the reference concentration $c_0=a/k$, which was fixed to unity in the simulations. Dropping tildes, the MIPS density dynamics reads $\partial_t\phi=\alpha_0\nabla\cdot\left[\phi\nabla(\mu_h(\phi)-\nabla^2\phi)-\mathrm{Pe}_C\phi\nabla c\right]$, whereas the diffusive KS dynamics reads $\partial_t\phi=\alpha_0\left[\nabla^2\phi-\mathrm{Pe}_C\nabla\cdot(\phi\nabla c)\right]$. Here, $\alpha_0=M_0/D_c$ and $\mathrm{Pe}_C=\chi_0/M_0$ for MIPS transport, while $\alpha_0=D_\phi/D_c$ and $\mathrm{Pe}_C=\chi_0/D_\phi$ for diffusive transport. These equations were coupled to $\partial_t c=\nabla^2c-\mathrm{Da}_0c+\mathrm{Da}_0\phi$ for minimal signaling or $\partial_t c=\nabla^2c-\mathrm{Da}_0\phi c+\mathrm{Da}_0\phi$ for autocrine signaling, with $\mathrm{Da}_0=\kappa k/D_c$. Details of the nondimensionalization and numerical implementation are provided in the Supplemental Material.\\

All simulations were performed on a uniform two-dimensional $100\times100$ grid with periodic boundary conditions. Spatial derivatives were discretized using second-order central finite differences. Time integration was performed using an implicit solver with adaptive time-stepping to ensure numerical stability. The cell-density field was initialized as a homogeneous state $\phi_0$ with small-amplitude Gaussian noise of standard deviation $0.02$, while the chemoattractant field was initialized at its corresponding homogeneous steady-state value $c_0=a/k$, which was fixed to unity throughout the simulations. Simulations were run until emerging patterns reached a quasi-steady morphology. Unless otherwise stated, we fixed $\phi_0=0.8$, $\phi_m=0.9$, $Pe_R=10^{-3}$, and $c_0=a/k=1$. These representative baseline values place the system in the MIPS-unstable regime while allowing systematic variation of $\mathrm{Pe}_C$, $\alpha_0$, and $\mathrm{Da}_0$.

\subsection*{Code availability} 
The source code developed for this study is publicly available on GitHub at \url{https://github.com/Endres-group/CMIPS_and_KS_paper}.

\subsection*{Acknowledgements}
We are thankful for useful discussions within the Physics of Life Network of Excellence at Imperial College London. We also acknowledge funding from the AI-4-EB Consortium for Bioengineered Cells and Systems (BBSRC award BB/W013770/1) (to RGE).

\bibliography{ref}

\end{document}